\documentclass{PoS}

\newcommand{\feoh}{\textrm{[Fe/H]}}

\newcommand{\xfe}{\textrm{[X/Fe]}}
\newcommand{\xh}{\textrm{[X/H]}}

\newcommand{\hyp}{\ \mathchar`-\ }

\newcommand{\loge}{\ensuremath{\log \epsilon}}
\newcommand{\teff}{\ensuremath{T_{\textrm {eff}}}}

\title{SAGA: Stellar Abundances for Galactic Archaeology}

\ShortTitle{SAGA: Stellar Abundances for Galactic Archaeology}

\author{\speaker{Takuma Suda}\\
        National Observatory of Japan, Osawa 2-21-1, Mitaka, Tokyo, 181-8588, Japan\\
        E-mail: \email{takuma.suda@nao.ac.jp}}

\abstract{A tutorial for the Stellar Abundances for Galactic Archaeology (SAGA) database is presented. This paper describes the outline of the database, reports the current status of the data compilation and known problems, and presents plans for future updates and extensions.}

\FullConference{XII International Symposium on Nuclei in the Cosmos,\\
		August 5-12, 2012\\
		Cairns, Australia}

\begin{document}

\section{What is the SAGA database?}
The SAGA database is an online tool for retrieving observational data of extremely metal-poor (EMP) stars\cite{Suda2008} (hereafter, Paper I)\footnote{http://saga.sci.hokudai.ac.jp or http://astro.keele.ac.uk/saga/}.
The database enables plotting of viewgraphs to show correlations between any combinations of quantities such as abundances, stellar parameters, and photometric data originally taken from the literature based on the observations of high-resolution spectroscopy\cite{Beers2005}.
Plotted data can be downloaded via web browsers together with figure files and a gnuplot\footnote{http://www.gnuplot.info} script to reproduce the last created figure.
Users can also upload files to compare the data on the browsers.
All the available quantities in the database are listed in Table 1 of Paper I.
This tool is helpful not only for looking up the available data but also for comparing models and observations, which can be applied to various astrophysical problems.
For example, abundances patterns of EMP stars can be compared with nucleosynthesis models of supernovae\cite{Tominaga2010}, AGB yields\cite{Stancliffe2009,Cristallo2009,Nishimura2009,Suda2010,Lugaro2012}, planetary nebulae\cite{Otsuka2010}, yields from rotating massive stars\cite{Meynet2010}, and so on.
Comparisons with Galactic chemical evolution models\cite{Komiya2010} and binary population models\cite{Izzard2009,Pols2012,Suda2012} are also possible.
We can obtain the characteristics of EMP stars by investigating the abundance trends, evolutionary status, and/or kinematic information.
Of course, usage of the database is not restricted to these examples, and users can find their own ways to use the data.
However, as we will discuss later, we need careful inspection of the data because the SAGA database is simply a collection of available data in the literature.

\section{Current status of the data}

\begin{figure} 
\includegraphics[width=.6\textwidth]{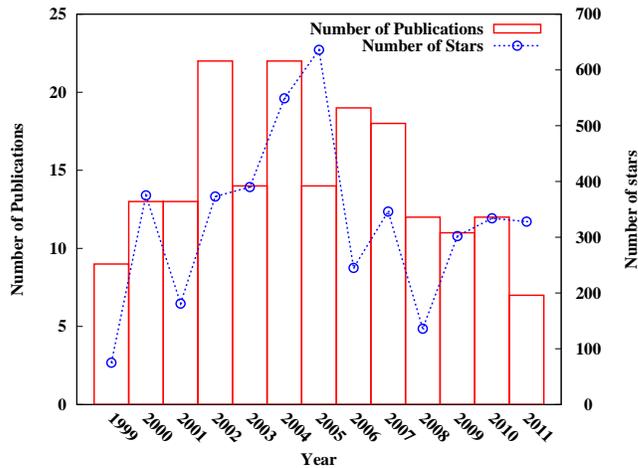} 
  \caption{The number of publications registered in the SAGA database.
           All the papers published by 1999 are counted as "1999". The oldest paper in the database was published in 1992.
          }
  \label{fig:pub}
\end{figure}

Figure~\ref{fig:pub} shows the number of compiled papers in the database in each year.
The current criteria for compilation are papers published since 2000 and those including stars with $\feoh \leq -2.5$.
The former criterion is related to the availability of the data in electronic format in most cases, but some papers published before 2000 were included.
If one of the target stars satisfies the latter criterion, then all the stars in the paper are included.
The figure also shows the number of stars registered in each year.
As seen in the figure, the number of stars registered in the database does not have any increasing or decreasing trends for the last decade, while the number of published papers dealing with EMP stars is slightly decreasing.
Note that there are several papers published in 2011 that are still to be compiled.
Therefore, roughly speaking, we have more than ten new papers on the observations of EMP stars every year.

As of August 12, 2012 (the version of August 6, 2012), the number of unique stars in the database is 1488 (4270 stars when independently reported measurements are included).
The number has increased by 276 since the report in Paper I.
The number of EMP stars ($\feoh \leq -2.5$) is 436 and has increased by 44 since the publication of the database.
The number of records for abundance data is more than 27000 for 62 species from lithium to uranium, which has significantly increased from $\sim 17000$ at the first release of the database.

\begin{figure} 
\includegraphics[width=.6\textwidth]{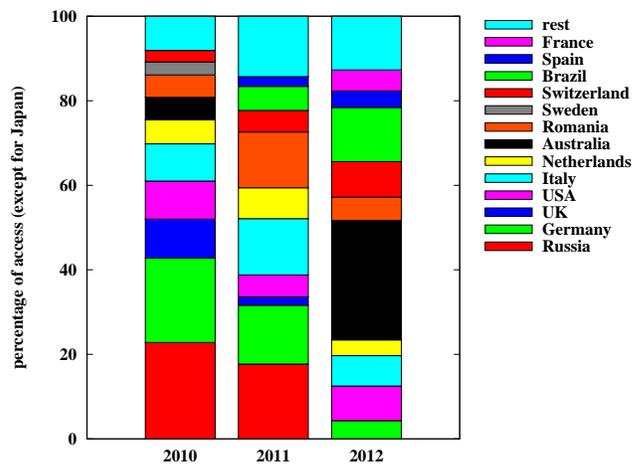} 
  \caption{Percentages of access from country to country. The data is based on the file access using the software {\it Webalizer} to record access logs % \footnote{ http://www.mrunix.net/webalizer/ }
  and has been available since 2010. The top ten countries are presented except for Japan due to the dominant fraction of access from a Japanese domain. About 70 to 80 per cent of the access to the database is from Japan.}
  \label{fig:access}
\end{figure}

Figure~\ref{fig:access} shows the percentages of access to the website of the SAGA database for most visited countries, except for access from Japan, since most of the access is from Japan ($\sim 70 \hyp 80$ \% of total access).
The data has been available since 2010 and is based on the file access using a software for access logs\footnote{Webalizer, http://www.mrunix.net/webalizer/ }.
Accesses from unknown nationality (in most cases, .com and .net, which makes a significant contribution) are excluded from the data.
The data in the figure apparently shows a correlation with the number of active research people/groups working on stellar models and observations.

% Figure: publication year

% Figure: access

\subsection{A new tool for data retrieval and plotting}

% Figure: viewer
\begin{figure} 
\includegraphics[height=.95\textheight]{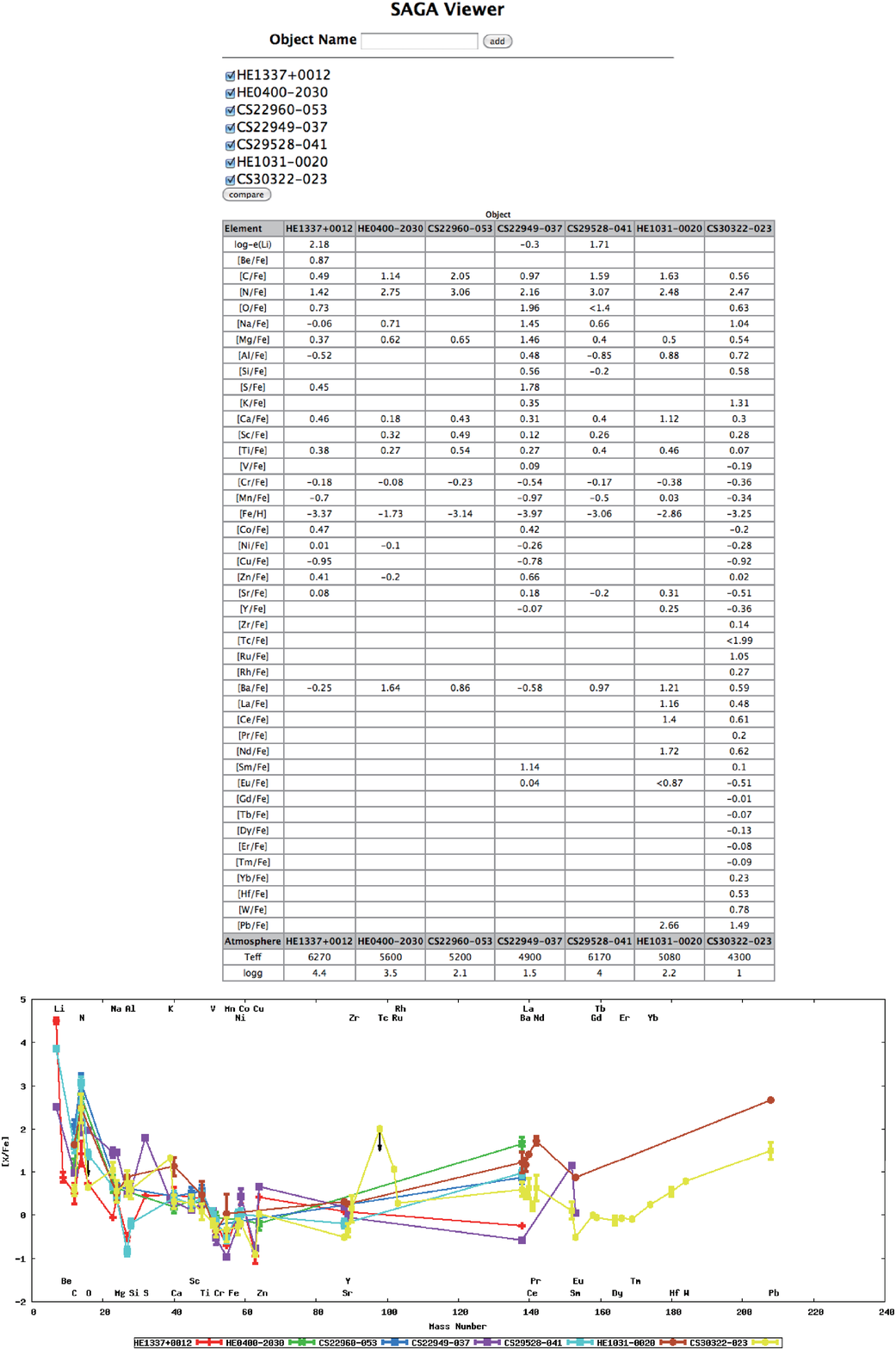} 
  \caption{Screenshot of the SAGA viewer.}
  \label{fig:viewer}
\end{figure}

Since the update in November 2010, another data retrieval sub-system has been available that enables users to view and compare the abundance patterns of individual stars.
The tool has temporarily been named "SAGA viewer" and is accessible from the top page of the SAGA database.
The screenshot is shown in Figure~\ref{fig:viewer}.
Using this new system, users can compare the abundance patterns for any stars in the database by plotting the abundances as a function of mass number.
The user interface is very simple in the current version:
(1) enter the name of the object with the help of the autocomplete function, and add as many as users like.
(2) press the "compare" button to see a data table and a summary viewgraph for selected stars.
For those element abundances with multiple sources of data, the values shown in the data table are chosen based on the same algorithm as the data retrieval sub-system.
The sources of the data can be traced by the mouse-over links on the adopted values in the data table.

\section{How to use the database}

Instructions for using the database appear on the SAGA database website.
The database and the user interface will be updated periodically, and therefore the users are recommended to check for updates on the site when they visit.
The tutorial for the data retrieval sub-system is linked from the top page.
When users bookmark the website, it is strongly recommended that they add the top page (wiki page) to their list of bookmarks, not the contents of the database such as the data retrieval sub-system and the SAGA viewer.
Otherwise, users will miss the important information about the updates of the database and the web system.

In the last update of the website, the "readme" page was added to the main contents.
It contains important notes when the database is used.
For example, it gives a description about the software to create figures, which will be helpful for managing downloaded data, and a script to reproduce the last figure in their local environments.
It also describes general rules in the database about the identification of the objects and the conversions of element abundances between different units such as $\log \epsilon$ and \xfe.
The conversion of units is important for homogeneous treatment of the data and is discussed in the next section.
The contents of the readme will be updated along with the future updates of the system.

If users find any problems in understanding how to use the database, they are encouraged to contact the author.

\section{Known problems in the database}

There are some problems that arise in the database when we treat the data as a homogeneous data set.
The data in the SAGA database cannot be homogeneous in the sense that the derived abundances are based on different facilities, different methods for analyses, different adopted parameters for stellar atmosphere, etc.
In other words, we cannot remove systematic uncertainties caused by these factors.
In spite of this difficulty, the SAGA database tries to remove the potential systematic differences among the literature.

The current version of the SAGA database adopts abundances based only on 1D LTE analyses.
The corrections to the final abundances by the NLTE and 3D analyses can be larger than typical uncertainties between multiple measurements ($\sim 0.3$ dex, Paper I, and a discussion below), which can be a major source of systematic differences between multiple measurements.
%Of course, it is not a good approximation to treat an actual stellar atmosphere as a plane parallel with the condition of LTE satisfied.
In particular, NLTE corrections are important in EMP stars due to the decrease of thermalised electrons in these stars, which gives rise to a departure from LTE \cite{Asplund2005a}.
Therefore, recent papers try to construct more realistic models for stellar atmospheres\cite{Asplund2009}.
The trend is obvious in the list of candidate papers for compilation in the database.
In particular, the number of papers reporting abundances using NLTE and 3D analyses has increased rapidly since 2009.
Among them, there are a few papers that report the abundances based on NLTE models only without publishing the results with LTE model atmosphere.
If this is the major trend in publishing data in future, we will not be able to deal with the data as they are, and may need to consider (NLTE) corrections to all the existing data.
%This is briefly discussed in the next section.

\begin{figure} 
\includegraphics[width=.95\textwidth]{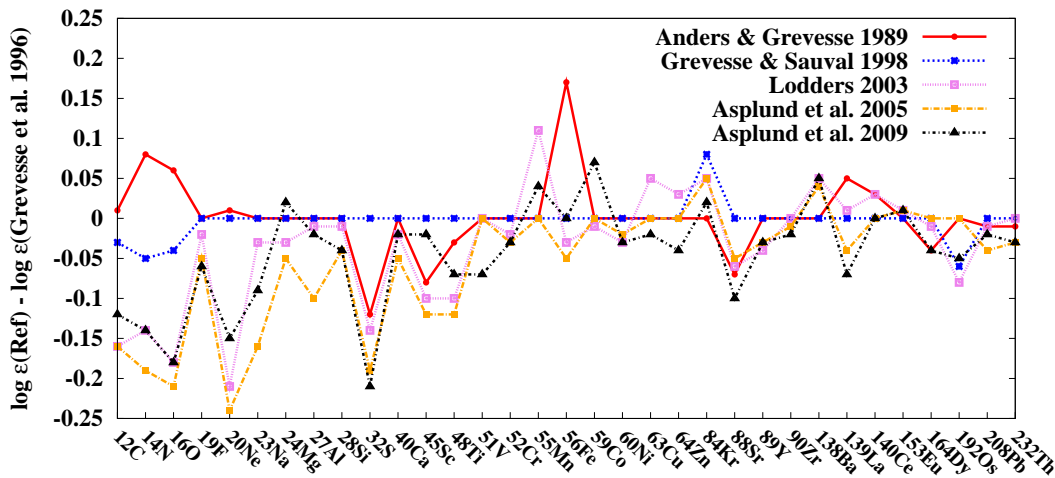} 
  \caption{Comparisons of the solar abundances by deviations from the abundances in Grevesse et al. (1996)\cite{Grevesse1996} from which the database adopts as reference values.
           In the top right corner, given are the comparison papers\cite{Asplund2009,Anders1989,Grevesse1998,Lodders2003,Asplund2005b}.
		   Abundance values are taken from photospheric abundances except for F, Os, Pb, and Th that are taken from meteoritic abundances.
  }
  \label{fig:sun}
\end{figure}

Another important problem is the difference in the adopted solar abundances.
In deriving the abundances in units of \xfe\ or \xh, we have to assume the solar chemical composition.
In most papers, the authors use the data from a set of the most up-to-date compilation of solar abundances in the literature.
Some authors take values from different papers for different elements, or measure the solar abundances by themselves.
These different normalisations cause a difference in abundances typically up to $\sim 0.1$ dex depending on elements.
The comparison of the major data sets of the solar abundances is shown in Figure~\ref{fig:sun}.
For most element species shown in the figure (for which abundances are derived in many papers), the differences are smaller than $0.1$ dex.

% Figure: solar abundances

In the current version of the database, abundances of element X are calculated by the relation, $\xfe = \xh - \feoh$ if both \feoh\ and \xh\ (or \xfe) are available.
Otherwise, the abundances are calculated from the value of $\log \epsilon$ using the solar values of Grevesse et al. (1996)\cite{Grevesse1996} with the relation $\xh = \log \epsilon(X) - \log \epsilon(X)_{\odot}$.
Accordingly, the solar abundances sometimes can be inconsistent within a paper if the original paper does not adopt the solar values from Grevesse et al. (1996).
However, this is not the case for most papers when we compare the values of \xfe\ and \xh\ because there are few papers presenting their results only in units of $\log \epsilon$.
In the case of comparisons between two or more papers by the values of \xfe\ or \xh, there exists a systematic difference unless the same values of the solar abundances are used.
%For the derivations of \xfe\ and/or \xh\ from $\log \epsilon$, we use Grevesse et al. (1996) with the relation $\xh = \log \epsilon(X) - \log \epsilon(X)_{\odot}$.
%This is not the case either for most papers because there are few papers presenting their results only in units of $\log \epsilon$.

\begin{figure} 
\includegraphics[width=.95\textwidth]{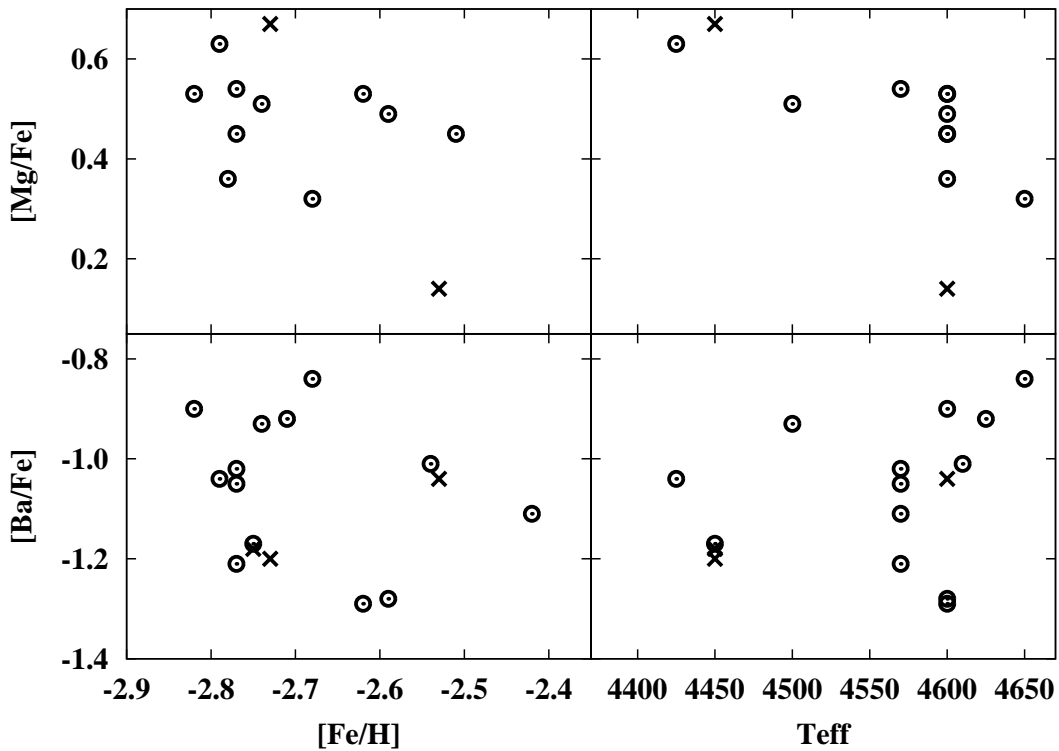} 
  \caption{Mg and Ba Abundances for HD122563 as functions of \feoh\ and \teff\ derived by different papers.
           The data points shown by crosses are derived from the abundance conversion using the solar values of Grevesse et al. (1996)\cite{Grevesse1996}.
  }
  \label{fig:bamg}
\end{figure}

It may be useful to compare the abundances of a specific object for which many papers report them.
As an example, all the Mg and Ba abundances for HD 122563 available in the database are plotted in Figure~\ref{fig:bamg}.
The star HD 122563 is a well-known reference star to calibrate abundances.
There are 34 papers measuring the abundances of this star, 17 of which report Mg and/or Ba abundances.
The data points shown by circles are taken from the literature without any corrections, while those shown by crosses are corrected values using our adopted solar value\cite{Grevesse1996} because these data are available only in \loge\ or \xh.
Note that there is an inconsistency in the adopted solar abundances as mentioned above.
Unfortunately, only 10 papers in the plot explicitly describe the adopted solar abundances, and hence, it is not clear how the effect of the correction with the solar abundances appears in the figure.

% Figure: HD122563

\section{Future plans for updates and extensions}

The SAGA database covers the data of the Galactic metal-poor halo stars with $\feoh \leq -2.5$ that are thought to comprise a minor population in the Galaxy.
Thanks to the great efforts by those working on the observations of halo stars, we can investigate the characteristics of extremely metal-poor stars with the growing number of such stars.
We have several significant outliers among stars at $\feoh \lesssim -2.5$ while there is a ``normal'' population\cite{Norris2012,Yong2012} that are thought to be the result of supernovae at the early phase of the Galactic chemical evolution.
However, we still need more data to understand the evolution of the Galaxy and the universe through {\bf the nuclei in the cosmos}.
As shown in Fig~\ref{fig:pub}, the number of observed stars and papers has not increased significantly due to the limitation by the analyses of high-resolution spectra, which are not done automatically in most cases.
On the other hand, the analyses can be performed with the SEGUE Stellar Parameter Pipeline (SSPP), which is a tool for automatically deriving stellar parameters\cite{Beers2012,Lee2008} and which will significantly enlarge the sample size of halo stars in the Galaxy in a couple of years.
Therefore, the database will be more and more important for future research.
Here we briefly describe our future plans.

We are planning to extend the database by removing the metallicity threshold.
Of course it is impossible to cover all the observed stars with $\feoh > -2.5$.
The number of stars increases by orders of magnitude compared with the known extremely metal-poor stars.
Therefore, it is difficult to construct a complete metallicity distribution function of the Galaxy by simply collecting data in the literature.
In addition, research on most of the observed stars with $\feoh \gtrsim -2$ is not necessarily published by those who search for the most metal-poor stars.
For these reasons, the compilation of metal-rich stars will be a minor update by adding several papers on the observations of metal-poor stars that were not compiled before because of the absence of stars with $\feoh \leq -2.5$.
Other papers dealing with Pop.~II stars (typically $\feoh > -2$) will be added to increase the number of stars and widen the coverage of metallicity in the database, although these samples will not be homogeneous nor necessarily appropriate for a statistical treatment.
We believe this extension will be of some help to compare models with observations.

Another plan is to construct a database for stars in dwarf spheroidal galaxies (T. Suda, J. Hidaka, W. Aoki et al., in prep.).
We have a list of papers dealing with these stars and have registered them using the reference management sub-system (Paper I).
Some of the papers have been compiled and are ready to use.
In a future update, we will be able to compare stars in both the Galaxy and the local group galaxies by plotting all the data in the same viewgraph.

In addition to increasing the number of data, we can improve the quality of the existing data.
As stated in the previous section, the database can be improved by more efforts to homogenise the data.
A possible way to do this is to reanalyse the data using the available data on equivalent width like Yong et al. (2012)\cite{Yong2012} did, although it is not our plan at the moment.
Another possibility is to complete the data for adopted solar abundances in the literature.
This is not a complete way, but the results can be shared with anyone and should be useful.
On the other hand, the treatment of stellar parameters is still an open question.
Analyses with 1D LTE were quite popular until the last several years, but non-LTE analyses and 3D effects in stellar atmospheres have been capturing the market for model atmospheres in recent years.
We should continue to watch trends in this research field and be ready for better treatment of data.

The project of the SAGA database finds another way to expand.
It started with the author's experience with the activity on nuclear data\footnote{Japanese Charged Particle Nuclear Reaction Data Group: http://www.jcprg.org/}, and the feedback goes to the field of nuclear physics again.
We started a project for the database of nuclear equation of state (EOSDB)\cite{Ishizuka2012}, which aims at applications to astrophysical problems by publishing many available nuclear equation of state.
It shares a common interface with that used in the data retrieval sub-system in the SAGA database as in the case of the search and plot system for cross sections of nuclear reaction rates, to the development of which the author contributed.

There are other possibilities for updates of the database.
Future observational projects using big telescopes such as GMT, E-ELT, and TMT, and large survey projects like Gaia will significantly change the quantity and quality of data.
It will be important for us to work out a better strategy to abstract information efficiently and to get a deeper insight into the Galactic archaeology with stellar data.

\acknowledgments
The author is grateful to Nicolas Grevesse for useful comments on the abundance analyses of extremely metal-poor stars.
The author thanks Yolande McLean for proofreading of the manuscript.
This work has been supported by Grant-in-Aid for Scientific Research (23224004), from the Japan Society of the Promotion of Science.

\end{document}